\documentclass[epj]{svjour}
\usepackage{graphics}
\begin{document}
\title{Aging phenomena in nonlinear dissipative chains: Application to polymer.}

\author{Franck Gobet, Sergio Ciliberto, Thierry Dauxois}
\offprints{Franck Gobet}          
\institute{Laboratoire de Physique, UMR-CNRS 5672, ENS Lyon, 46
All\'{e}e d'Italie, 69364 Lyon c\'{e}dex 07, France}
\date{Received: \today / Revised version: day}
%
\abstract{ We study energy relaxation in a phenomenological model
for polymer built from rheological considerations: a one
dimensional nonlinear lattice with dissipative couplings. These couplings
are well known in polymer's community to be possibly
responsible of $\beta$-relaxation (as in a Burger's model). After
thermalisation of this system, the extremities of the chain are
put in contact with a zero-temperature reservoir, showing the
existence of surprising quasi-stationary states with non zero
energy  when the dissipative coupling is high. This strange
behavior, due to long-lived nonlinear localized modes, induces
stretched exponential laws. Furthermore, we observe a strong dependence
on the waiting time t$_w$ after the quench of the two-time
intermediate correlation function C(t$_w$+t, t$_w$). This function
can be scaled onto a master curve, similar to the case of spin or
Lennard-Jones glasses.
} 
\authorrunning{Gobet, Ciliberto and Dauxois}
\titlerunning{Aging phenomena in nonlinear dissipative chains}
\maketitle
\section{Introduction}

An important challenge in polymers physics nowadays is the
understanding of non-equilibrium glassy state and associated slow
relaxations \cite{ediger}. At temperature below the
liquid-to-glass transition temperature, the structural relaxation
time depends indeed on the time spent in the glassy phase, the
so-called waiting time $t_w$: this is the aging effect
\cite{struck,young,bellon}. Low-frequency mechanical
spectroscopies show two different slow relaxations, namely
$\alpha$ and $\beta$, corresponding to two different temperatures
T$_\alpha> $ T$_\beta$ which depend on the mechanical excitation
frequency; T$_\alpha$ is shown to be very close to the
liquid-to-glass transition temperature \cite{ediger}. Both
relaxation modes are characterized by a maximum of dissipation but
are surprisingly differently affected by physical aging: strongly
for the first one but very weakly  for the second. In spite of
many studies, the microscopic origin of liquid-to-glass transition
and of these relaxation modes are not yet well established
\cite{conf}. The $\alpha$~relaxation, associated to the
liquid-to-glass transition, involves cooperative motions inducing
an increase of the viscosity, whereas localized motions would be
responsible of the $\beta$ relaxation  with a clamping of degrees
of freedom (for instance macromolecule rotation) near the
temperature~T$_\beta$.

\begin{figure}[h]
\centerline{\resizebox{8.6cm}{9.17cm}{\includegraphics{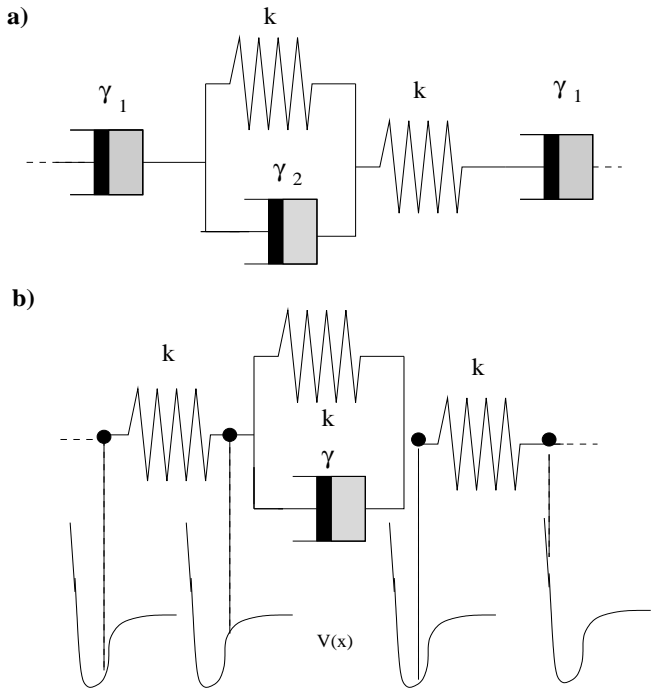}}}
\caption{Sketch of the Burger's linear model (panel~a), and of our
non-linear dissipative model for polymers (panel~b).}
\label{burger}
\end{figure}

Some phenomenological models (for instance Burger, Maxwell,
Kelvin-Voight or Zener's models) have been developed using linear
viscoelastic theory~\cite{ferry,eirich} and describe easily some
aspects of relaxation in spite of their simplicity. These systems
(see for example Burger's model in Fig. \ref{burger}a) correspond
to a succession  of springs and viscous elements (dashpots) set in
parallel or in series, to modelize interactions between and inside
polymers. As we will see in Section 2, in the case of Burger's
system, the dashpot with viscous parameter $\gamma_1$ (which
modelized viscous interaction between molecules) is linked to
$\alpha$ relaxation (and therefore to liquid-to-glass transition),
whereas the second piston (which modelized a dissipative
interaction inside the molecules) induces $\beta$ relaxation. Let
us notice that these simple dynamical models don't exhibit any
physical aging. One way to describe and explain this phenomenon
could be that of taking into account nonlinear interactions inside
macromolecules.

Indeed, one of the most important feature of nonlinear systems is
the localization of vibrational energy (named discrete breather)
that modifies strongly the energy relaxation \cite{aubry,reigada}
and induces non-equilibrium dynamics. Brea-thers are known to have
very interesting dynamical properties and have been used in
explaining a variety of physical and biophysical phenomena. For
instance in nonlinear systems where breathers become mobiles, they
could contribute directly to the energy transfer and modify its
relaxation properties in a nonexponential dependance. This
interesting phenomenon has been invoked in several physical settings
as DNA molecules \cite{peyrard}, hydrocarbon structures
\cite{kopidakis}, targeted energy transfer between donors and
acceptors in biomolecules \cite{aubry2}. When the coupling is much
smaller than the non linearity, the presence of essentially pinned
long-lived breathers in nonlinear systems blocks the energy
propagation \cite{aubry}. The macroscopic manifestation of this
phenomenon is a very slow relaxation of the total energy,
reminiscent of the long lifetime of metastable states in glassy
systems observed after a quench.

The case of relaxation phenomena in nonlinear lattices
with bulk dissipation (as in glassy polymers)
has received much less attention, though the experimental systems
belong to this class \cite{marin}. Indeed, the $\beta$ relaxation
in glassy polymers, which is characterized by a maximum of
dissipation, may be induced by the clamping of degrees of freedom.
In spite of this dissipative effect, physical aging of this system
is very slow. Understand this strange behavior is one of the aims
of this paper. To accomplish this goal, we will examine the
influence in  energy relaxation of a dissipative coupling by
performing numerical studies on a phenomenological model of
polymers. The system, described in Section 3 and pictured in
Fig.~\ref{burger}b, corresponds to  particles coupled via elastic
and dissipative interactions (as in Burger's model for the
molecular modelization); in addition, each particle is submitted
to an on-site nonlinear potential to take into account
interactions between different polymers including several
contributions such as hydrogen bonds linking and repulsion of two
chains (short-range steric restrictions). For a fixed viscous
parameter $\gamma$, we examine the relaxation of energy when,
after thermalization, the ends of the chain are placed in contact
with a zero-temperature reservoir. Results show different kinds of
energy relaxation regime which depend strongly of the dissipative
terms: in particular, we show that the system can relax very
slowly in spite of high dissipative couplings!

\section{Viscoelastic systems and $\alpha$, $\beta$ relaxations}

In many experimental studies, isochronal dynamical mechanical
spectrometry has been performed at different frequencies
\cite{ediger}. With these measures, it is possible to map the loci
of $\alpha$ and $\beta$ relaxations (defined by maximum of the
imaginary parts of shear modulus G=G'+iG'') versus temperature and
frequency. Typically, the $\alpha$-relaxation process exhibits a
clearly non-Arrhenius behavior above the liquid-to-glass
transition T$_g$, whereas an Arrhenius behavior is observed for
temperature below T$_g$, when the system is out of thermodynamical
equilibrium. The second process $\beta$ appears at higher
frequencies than $\alpha$-process with also an Arrhenius behavior.

It is well-known in polymer's community that
some aspects of the rheology of
glassy polymers can be described easily by using phenomenological
models from viscoelastic linear theory~\cite{ferry,eirich,turner}. 
It can be assumed that the
deformation of the polymer is divided into elastic and viscous
components and can be described by a combination of Hooke and
Newton's laws. These systems correspond to
springs and dashpots, either in series or in parallel. The
Burger's model (one element is described in Fig.~\ref{burger}a)
seems to be a good candidate to illustrate the rheological
behavior of polymers in glassy state. One element is made of two
springs in series to modelize  the chain of a macromolecule. In
addition, one part of this chain is coupled in parallel with a
dashpot to take into account the clamping of some degrees of
freedom at low temperature, inducing the $\beta$-relaxation. The
second dashpot in series with elastics components illustrates the
viscous behavior of the system when it becomes glassy and induces
the $\alpha$-relaxation.

In this model, the shear modulus G=G'+iG'' is measured by
applying a sinuso\"{\i}dal force, with frequency $\omega$, at one
end of a chain with 100 elements and by fixing the other
extremity. G is the ratio between the applied force
and the relative displacement of the chain extremity obtained in the stationary limit. 
The shear modulus is determined in a temperature window
by using two Arrhenius laws for the viscosity $\gamma_1$ and
$\gamma_2$:
\begin{eqnarray}\gamma_1=\gamma_{1}^0\ e^{E/k_bT}
\end{eqnarray}
and $\gamma_2=10^{-5}\gamma_1$. 
Furthermore, as in all viscoelastic linear studies 
(see for instance references ~\cite{ferry,eirich,turner})
there is no inerty inside the system: each connection between springs and dashpots
is considered as a particule with a mass equal to zero.

\begin{figure}[h]
\centerline{\rotatebox
{270}{\resizebox{6.39cm}{8.6cm}{\includegraphics{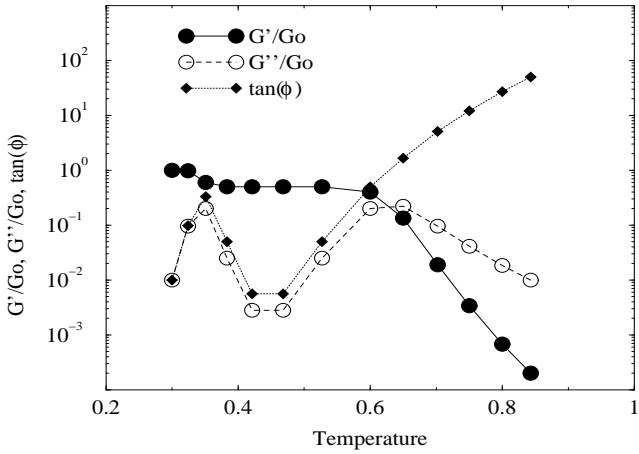}}}}
\caption{Values of the real (G') and imaginary (G'') parts of
shear modulus and of $\tan\phi$; they are plotted 
versus temperature for the Burger's model of Fig.~\ref{burger}a.
The shear moduli are divided by the modulus limit of low
temperature G$_0$. Parameters used in the calculations are k=0.1,
$\gamma_{1}^0$=10$^6$, E/k$_b$=9.7 and $ \omega$=10$^{-2}$.} \label{response}
\end{figure}

In Fig.~\ref{response}, we report
the evolution with temperature of the shear moduli G', G'' and
loss tangent $\tan \phi$=G''/G' for parameters given in the corresponding caption. 
These curves show two distinct regions of
rheology, namely for increasing temperature, the $\beta$-process
(weak decreasing of G' and, maxima of G'' and $\tan\phi$) and the
$\alpha$-process (strong decreasing of G' and maximum of G'').
This behavior is qualitatively similar to those observed
experimentally \cite{ferry}. However, some differences appear. The
loss tangent   exhibits an increase with temperature for the
$\alpha$-process, but we don't observe a maximum unlike in
experimental measurements. Furthermore, for temperature higher
than the liquid-to-glass transition temperature, a non-Arrhenius
behavior is observed experimentally for the $\alpha$-relaxation.
This characteristic is not take into account in the evolution of
$\gamma_1$.

In spite of these differences, the major overriding conclusion for
this section is the possibility to observe rheological
characteristics, like $\beta$-process, of the glassy state by
considering a viscous coupling inside the molecule (linked to the
rotation of terminal groups or other side chain motions
\cite{ferry}). This modelization is however very concise and it is
not  possible to describe any aging phenomena because of the
linearity of components.

In the next part, we will describe a phenomenological model
including  dissipative couplings and nonlinear  interactions that
may modelize more precisely one chain of a macromolecule. The main
question we would like to address is the possibility to observe a
long lived non-equilibrium state in nonlinear systems despite
dissipative terms. In order to solve this problem, we simplify the 
system: the initial thermalized  nonlinear system is put in a
zero-temperature bath with a fixed {\em viscous parameter 
independent of temperature}.

\section{The phenomenological dissipative nonlinear model}

We consider a one dimensional chain of N=200 anharmonic
oscillators, with a nonlinear on-site potential V(x), with free
ends and nearest-neighbor elastic coupling potential (the coupling
being k). The sketch of the chain is reported in Fig.~1b.  In
analogy with Burger's model and in order to modelize the clamping
degree of freedom, we consider for one half of nearest-neighbors a
dissipative coupling ($\gamma$ is the dissipative parameter) in
parallel with elastic coupling. For the on-site potential V(x),
describing the interactions between two molecules (hydrogen
long-range attraction and steric short-range repulsion), we have
chosen  the Morse potential
\begin{equation}
        \mathrm{V(x)=\frac{1}{2}[1-e^{-x}]^{2}}
\end{equation}
which has the appropriate shape to describe the strong repulsion
when the chains are pushed toward each other (x$<$0) and the
vanishing interaction when the chains are pulled very far apart
(x$\gg$1). Each end oscillators of our chain can be also submitted
to an additional damping force. The equations of motion of this
chain are given in dimensionless form by:
\begin{eqnarray}
        \mathrm{\ddot{x}_{2i}}&=&\mathrm{-\frac{\partial V}{\partial x_{2i}}-k(2x_{2i}-x_{2i+1}-x_{2i-1})} \nonumber\\
&&\mathrm{-\gamma (\dot{x}_{2i}-\dot{x}_{2i-1})-\gamma '\dot{x}_{2i} \delta(2i-200)}
\end{eqnarray}
and
\begin{eqnarray}
        \mathrm{\ddot{x}_{2i+1}}&=&\mathrm{-\frac{\partial V}{\partial x_{2i+1}}-k(2x_{2i+1}-x_{2i}-x_{2i+2})} \nonumber\\
&&\mathrm{-\gamma (\dot{x}_{2i+1}-\dot{x}_{2i+2})-\gamma '\dot{x}_{2i+1} \delta(i)}
\end{eqnarray}
where x$_n$ is the dimensionless displacement of the nth
oscillator from equilibrium, $\dot{x}_n$ its velocity and $\delta$
the Kronecker delta function. The mass of the oscillators is set
to unity by appropriately renormalizing time units. The equations
of motion have been integrated using a fourth order Runge-Kutta
method.

To study energy relaxation, we consider k=0.01 and we initially
thermalize the system at temperature T=1 by using 
Nos$\mathrm{\acute{e}}$-Hoover thermostats \cite{nose}. This
temperature is much higher than the critical temperature
(T$_c$=0.2) of the "order-disorder" transition which characterizes
the non dissipative model $\gamma$=0 (for more detail see
references \cite{theo,dau}). Then, in average, the kinetic energy per
site is higher than the depth of Morse potential (which is equal
to 0.5 in our arbitrary units) and  we can consider our system as
in an initial "liquid" state: all the particules are in the
plateau of the Morse potential. In other words, there are no
interaction between chains of molecules. The thermalization
procedure is performed with $\gamma$=$\gamma$'=0 by using a chain
of three thermostats to provide  a good exploration of the phase
space \cite{nose}.

After thermalization, the connection with the heat bath is
turned off and the lattice is connected to a zero temperature
reservoir via the damping term with $\gamma$'=0.1. However, we
explore the interval [0,100] for the dissipative parameter
$\gamma$. As our purpose is to search for long lived
non-equilibrium state in spite of dissipative couplings, we prefer
to simplify the problem and not to consider a temperature
dependence of $\gamma$ like in the Burger's model of section~2.
The moment of connection with the zero temperature reservoir is
chosen as the origin of time. At each step of the integration of
equations~(2) and~(3), we evaluate the total lattice energy:
\begin{equation}
\mathrm{E=\sum_{i=1}^{200}\left[\frac{1}{2}
{\dot{x}_i}^2+V(x_i)\right]+\sum_{i=1}^{199}\frac{k}{2}(
x_{i+1}-x_{i})^2}
\end{equation}
and consider the symmetrized local energy per site:
\begin{equation}
        \mathrm{E_i=\frac{1}{2}{\dot{x}_i}^2+V(x_i)+ \frac{k}{4}(x_{i-1}-x_i)^2
+\frac{k}{4}(x_i-x_{i+1})^2}\quad.
\end{equation}
Total energy is expected to decrease with time and converge to a
zero value of the "frozen" state at equilibrium.

\section{Relaxation of the thermalized system}

\begin{figure}[h]
\centerline{\resizebox{8.6cm}{11.3cm}{\includegraphics{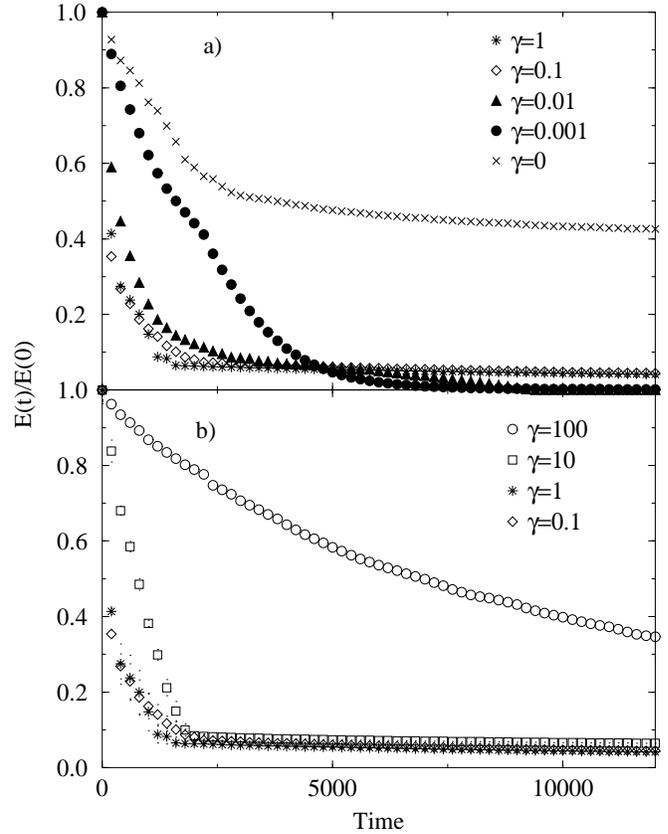}}}
\caption{Normalized total energy decay as function of time for various viscous parameter
$\gamma$. Initially each array is in thermal equilibrium at the temperature
T=1.}
\label{energie}
\end{figure}

In Fig.~\ref{energie}, we report the total lattice energy divided
by the initial energy versus time for various viscous parameter
$\gamma$. For $\gamma$=0, a clear non exponential decreasing
energy is observed, consistent with Tsironis and Aubry's
results~\cite{aubry}. This long-tail relaxation behavior was shown
by these authors to be connected to the presence of long-lived non
linear localized modes that are relatively mobile. If we consider
a small dissipative coupling ($\gamma$=10$^{-3}$) inside the
lattice, we see that total energy decreases faster than
previously. This small dissipative coupling induces a strong
modification of energy relaxation to the "frozen" state
(normalized energy closed to 0) for a time smaller than 10$^4$ (in
arbitrary units). This emphasizes that the dissipative couplings
change strongly the relaxation mechanisms and can induce a fast
relaxation regime.

For $\gamma$ higher than 0.1, a new surprising feature is
observed: normalized energy seems to be blocked with a very slow
decrease for long time whereas the dissipative parameter is
higher! The system seems to evolve in a quasi-stationary state
that is neither a "frozen" state (the normalized energy is clearly
different from 0) nor a "liquid" state" (energy is too low). We
can also notice that, at a given time, the energy of this
quasi-stationary state increases with $\gamma$. This situation
seems to be qualitatively similar to polymer systems where a very
long-lived non-equilibrium glassy state is observed with very slow
physical aging for temperature smaller than T$_\alpha$ (the
viscous parameter which follows an Arrhenius law is higher for
lower temperature).

\begin{figure}[h]
\centerline{{\resizebox{8.6cm}{8.6cm}{\includegraphics{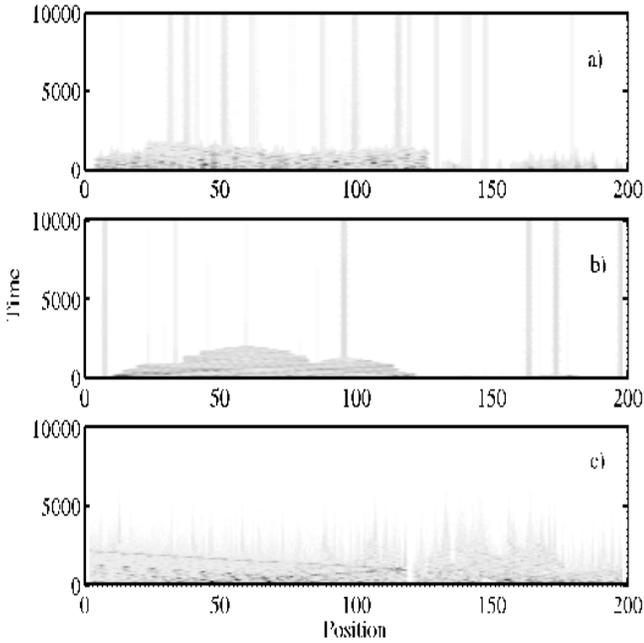}}}}
\caption{Evolution of the local energy E$_i$ along the chain for
various viscous parameter $\gamma$=10 (panel a), 0.1 (panel b) and
0.001 (panel c). The grey scale goes from E$_i$=0 (white) to the
maximum E$_i$-value (black).} \label{energiegamma}
\end{figure}

In previous relaxation studies in non linear lattices where
blocking energy was observed, it has been shown that such behavior
may be induced by long-lived breathers \cite{aubry}. In order to
examine more precisely the present situation, we report in
Fig.~\ref{energiegamma} the spatiotemporal energy landscape of the
lattice by plotting the local energy E$_i$ in each lattice site
for $\gamma=10^{-3},10^{-1}$ and 10. Time advances along the y
axis until t=10$^4$ and a gray scale is used to represent the
local energy with darker shading corresponding to more energetic
regions. For $\gamma$=10$^{-3}$, we can see two kinds of energy
relaxation: on the one hand, there is a dissipation of the energy
inside the lattice, characterized by a "fibrous structure" of the
local energy landscape in Fig.~\ref{energiegamma}c and, on the
other hand, we observe a dissipation of mobile breather via
surface damping characterized by "dark oblique lines". 
For $\gamma\geq$0.1, we notice the clear presence
of pinned long-lived breathers, responsible of the energy
relaxation blocking. This localisation of energy is observed after
a short time where energy not only decreases via the surface
damping but also via a dissipation inside the lattice (see for
example the landscape for $\gamma=10^{-1}$ and t$<$2500).
Furthermore, these states have a very long lived time and are still
observed for t higher than~10$^5$.

\begin{figure}[h]
\centerline{\rotatebox
{270}{\resizebox{6.75cm}{8.6cm}{\includegraphics{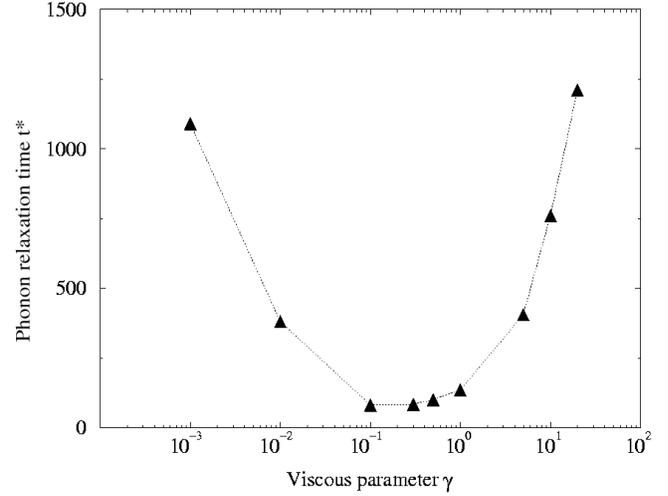}}}}
\caption{Phonon relaxation time t$^*$, defined by
E(t$^*$)/E(0)=0.5, for various values of the viscous
parameter~$\gamma$.} \label{phononsrelax}
\end{figure}

We have mentioned previously that, at early times, pho\-nons and
mobile breathers dissipation takes place before pinned breathers
relaxation. Typically the hierarchy of relaxations processes may
be classify in a sequence of characteristic times
t$_1^*<$t$_2^*<$... \cite{tsang}, where the energy relaxation
corresponds approximately to exponential or a stretched
exponential decay~\cite{bibaki}. Let us introduce the phonon
relaxation time t$^*$ defined by E(t$^*$)/E(0)=0.5 which is
characteristic of the phonon and mobile breather 
dissipation inside the lattice. We
have reported in Fig.~\ref{phononsrelax} the evolution of this
relaxation time versus the viscous parameter $\gamma$. We clearly
see, in this first step, that energy decreases faster for $\gamma$
between 0.1 and 1. In the case of a linear lattice, the maximum of
dissipation is predicted to correspond to a value of $\gamma$
which verifies $\gamma\omega \sim K+k$ where K corresponds to the
coupling constant of the linearized Morse potential (K=1 in our
case) and $\omega$ is the frequency of the phonon band that starts
at the frequency $\mathrm{\sqrt{K}}$=1 and extends to
$\mathrm{\sqrt{K+4k}}$$\sim$1.02. Therefore, in the case of linearized
oscillator, we expect a maximum of dissipation for $\gamma$ close
to 1 in agreement with what is reported in
Fig.~\ref{phononsrelax}. It is thus very surprising to observe
quasi-stationary states in a second regime whereas dissipative
effects are important. We have to notice that this paradoxal
effect is very similar to the situation observed in the case of
$\beta$ relaxation of polymers in glassy state: a very slow
physical aging in spite of the maximum of dissipation~\cite{conf}.

\begin{figure}[h]
\centerline{{\resizebox{8.6cm}{7.95cm}{\includegraphics{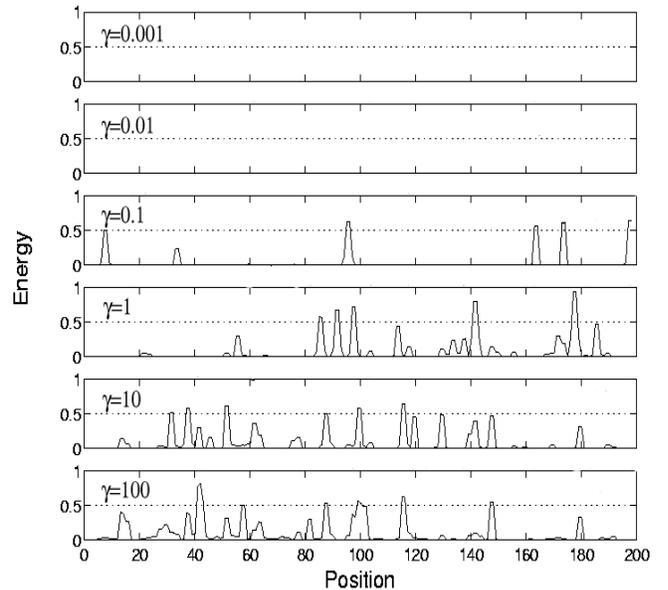}}}}
\caption{Instantaneous local energy E$_i$ along the chain at time
t=2.$10^4$ for various dissipative parameter $\gamma$. The dashed
lines show the limit of Morse potential for large displacement
x$_i$.} \label{energiesite}
\end{figure}

In Fig.~\ref{energiesite}, we report the local energy in the
lattice for various parameter $\gamma$ and at a given time
t=2.$10^4$ much higher than the characteristic time of phonons and
mobile breathers dissipation. For $\gamma$$<$0.1, the local energy
per site is close to 0 as seen previously in Fig.~\ref{energie}:
the system with small dissipative coupling reached thus rapidly an
equilibrium frozen state. For $\gamma$$\geq$0.1, we clearly see the
localization of energy, as long-lived breathers, in two
nearest-neighbors oscillators dispatched in some areas of the
lattice. We have verified that the two considered
nearest-neighbors are coupled via piston: when $\gamma$ is high
enough this coupling is clamped. In fact, the phase displacement
of velocities of the both particles is very small, inducing a very
slow dissipation and then referred a very long non-equilibrium
state. The energy of these long-lived breathers is close to 0.5 at
time t=2.$10^4$, and the associated sites are not linked with the
on-site Morse potential. This quasi-stationary state is clearly
not a frozen state, since some parts of a molecular chain are
"hot" but do not interact with other ones.

\begin{figure}[h]
\centerline{{\resizebox{8.6cm}{10.67cm}{\includegraphics{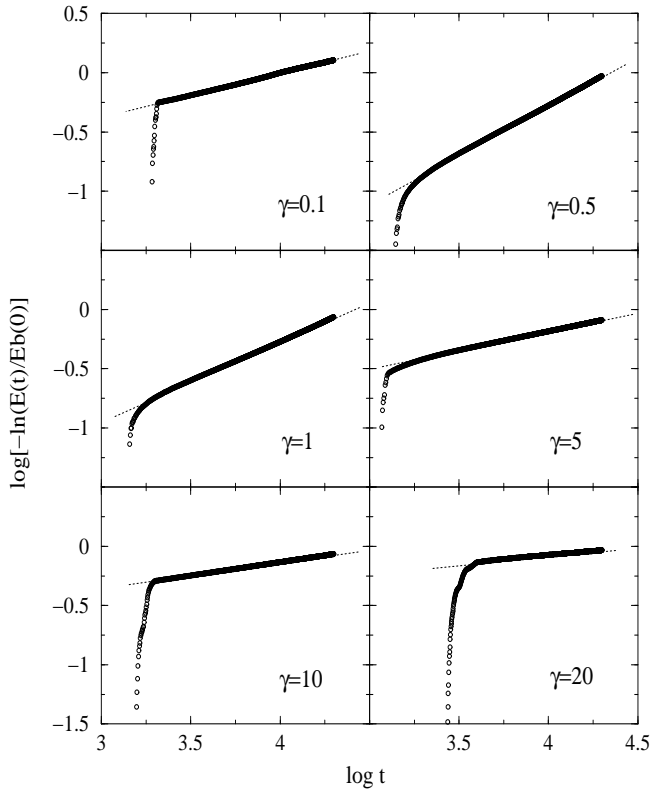}}}}
\caption{Log[-ln(E(t)/E$_b$(t))] versus log t for various
dissipative parameter $\gamma$. The dashed lines are the stretched
exponential fits.} \label{exponent}
\end{figure}

For t$\gg$t$^*$ and $\gamma$$\geq$0.1, the energy relaxation of the
quasi-stationary state can be very well fitted by a
Kohl\-rausch-Williams-Watts function or stretched
exponential function:
\begin{equation}
        \mathrm{E(t)=E_b(0)e^{-(a t)^b}}
\end{equation}
where the coefficients a, b and E$\mathrm{_b}$(0) are $\gamma$
dependent. E$\mathrm{_b}$(0) can be qualified as the total energy
of pinned brea\-thers at time t=0. The
Kohlrausch~\cite{kohlrausch} exponent b
 is a parameter measuring the deviation from a
single exponential form (0$\le$b$\le$1). In
Fig.~\ref{exponent}, we have reported the evolution of
[-lnE(t)/E$\mathrm{_b}$(0)] versus  t with logarithmic scales for
various parameter $\gamma$$\geq$0.1. We can see a very clear linear
dependance for high~t (after phonon and mobile breather
dissipation), which attests that this long-lived breather energy
relaxes as a stretched exponential.

\begin{figure}[h]
\centerline{\rotatebox
{270}{\resizebox{6.75cm}{8.6cm}{\includegraphics{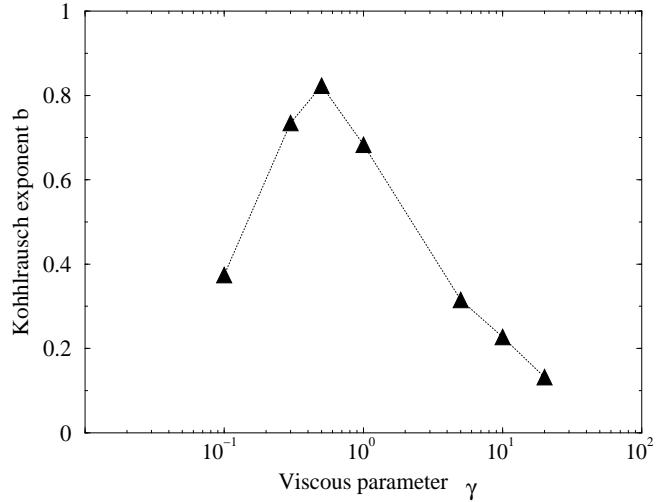}}}}
\caption{Kohlrausch exponent b for various viscous
parameter~$\gamma$.} \label{temps}
\end{figure}

We see also in these figures that the slope of the straight lines
depends on the viscous parameter $\gamma$. In order to examine
more precisely this dependence, we report in Fig.~\ref{temps} the
Kohlrausch exponent b versus the viscous parameter~$\gamma$. We
clearly see a maximum of the exponent b for values of~$\gamma$
close to 0.5. The b-value is equal to 0.82 and then the energy
relaxation differs from a pure exponential decay. Furthermore,
this figure shows that the pinned breathers relaxation is slower
for higher parameter $\gamma$. It is therefore  qualitatively
similar to the glassy state where a slower aging phenomena is
observed for slower temperature (and then higher  viscosity as we
can see in the Burger's model of section 2) \cite{struck}.

\section{Out of equilibrium dynamic correlations}

Studies of nonequilibrium systems like spin, structural or Lennard-Jones glasses
\cite{young,kob}
have shown that the nonequilibrium dynamics of the previously described states 
could be much efficiently characterized by two-time correlation 
functions of the form:
\begin{eqnarray}
\mathrm{C(t_w+t,t_w)}&=&
\mathrm{\langle A(t_w+t)A(t_w)\rangle} \nonumber\\
&&\mathrm{-\langle A(t_w+t)\rangle \langle A(t_w)\rangle}
\end{eqnarray}
where A is a microscopic observable, and t$_w$ is the "waiting time"
 i.e., the time elapsed after the quench. Brackets denote an 
average over different initial configurations at temperature T.
At equilibrium, this two-time quantity satisfies time translation
invariance  and then depends only on the time t. On the other hand, 
in out of equilibrium situations, such equilibrium property is
not verified: this function depends on the waiting time 
t$_w$ ("aging effect").  The correlations functions for large times
are expected to scale in the form:
\begin{eqnarray}
\mathrm{C(t_w+t,t_w)}&=&
\mathrm{C_{ST}(t)+C_{AG}\left(\frac{\xi(t_w+t)}{\xi(t_w)}\right).\label{eqncorr}}
\end{eqnarray}
The first term describes short time  dynamics that does not
depend on t$_w$ and has the equilibrium form.
The second term, or aging part, depends only on the ratio $\xi$(t$_w$+t)/
$\xi$(t$_w$) where $\xi$(t) is a monotonous increasing function of t.
In a lot of cases $\xi$(t)$\propto$t or $\propto$t$^\nu$ so that the aging 
part is simply a function of t/t$_w$ and exhibits a master curve (see for instance, 
experiments on 
thermoremanent magnetization \cite{vincent}, gels \cite{cipelletti}
or particle suspensions \cite{kroon,knaebel}).

In this study we have considered the microscopic observable 
A(t)=$\mathrm{(\sum^{200}_{i=1}x_i(t))/200}$ that is the
 mean deformation per site of the chain at time t. Numerical 
calculations have been done in the case of a quench of
the system with a viscous parameter $\gamma$=10 from temperature T=1 to T=0. 
Two-time correlation functions are obtained for various waiting time by 
considering 11 different initial configurations. Furthermore, in
order to make a quantitative comparison, we prefer to calculate correlation
normalized by C(t$_w$,t$_w$).

\begin{figure}[h]
\centerline{\rotatebox
{270}{\resizebox{6.75cm}{8.6cm}{\includegraphics{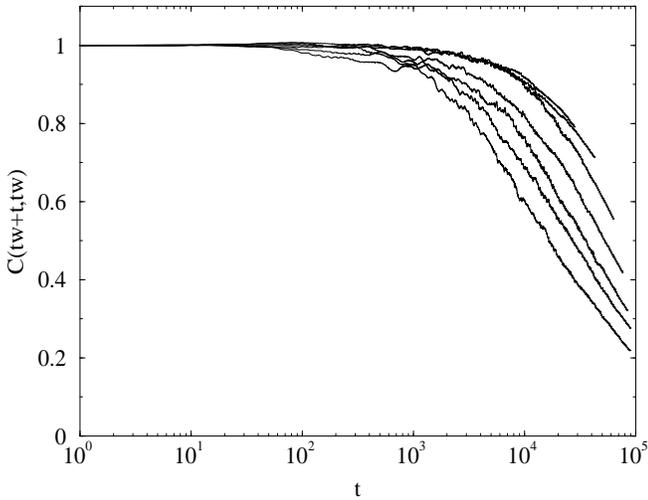}}}}
\caption{Normalized non-equilibrium correlation function C(t$_w$+t,t$_w$) 
for fixed value t$_w$ versus time t. 
The viscous parameter is $\gamma$=10 and from bottom to top
the waiting times are t$_w$=5844, 9300, 14700, 23137, 36321, 56920,
71225.} \label{correlation}
\end{figure}

In Fig.~\ref{correlation}, we have reported the evolution 
of normalized two-time correlations function
C(t$_w$+t,t$_w$) versus time t for different 
waiting time (we consider t$_w$$\gg$t$^*$ in order to study 
only long lived non 
equilibrium states after phonons and mobile breathers dissipation). 
The behavior of C clearly emphasizes the lost of 
time-translation invariance and 
the dependence on the waiting time t$_w$. This
 figure also shows that 
the dynamics can be decomposed into two time scales:

(i) at short time separation (t$<$20) 
correlation function doesn't depend of t$_w$ and is
 equal to the value expected 
at equilibrium (C=1 at T=0). 

(ii) the decay from this 
value toward zero arises
 in a second time scale that clearly depends on t$_w$: the 
system does'nt reach equilibrium
within the time window explored in the simulation.
Furthermore we can notice that the larger the waiting time, 
the longer it takes the system
to forget the configuration at time t$_w$. 
This behavior is typical of aging effect \cite{young}.

\begin{figure}[h]
\centerline{\rotatebox
{270}{\resizebox{6.75cm}{8.6cm}{\includegraphics{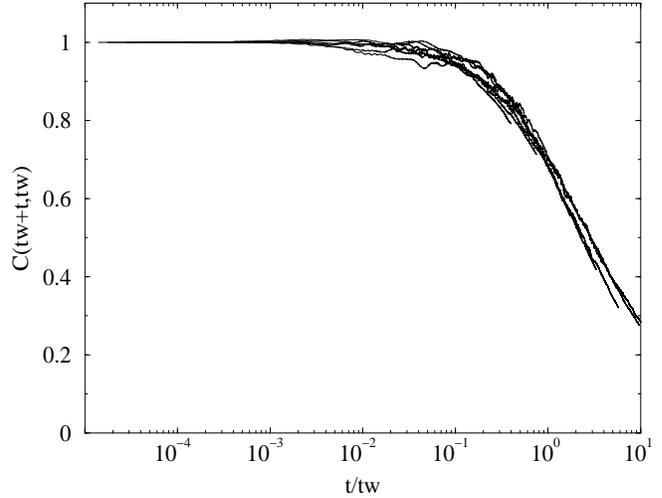}}}}
\caption{The data of Fig \ref{correlation} versus t/t$_w$: master curve}
 \label{maitresse}
\end{figure}

Guided by equation (\ref{eqncorr}) we test the scaling assumption for  long times 
in Fig. \ref{maitresse}, where normalized correlation functions are reported
versus normalized time t/t$_w$. The different curves can be superimposed, indicating the
validity of the scaling ansatz and the existence of a master curve.
This striking feature is observed in many non-equilibrium systems 
\cite{vincent,cipelletti,kroon,knaebel} 
and
is very similar with comparable studies 
on spin glasses \cite{young} or Lennard-Jones glasses \cite{kob}. 
The physical origin of this universal t/t$_w$ scaling is, at this day,
an open question. Kob and Barrat suggest in reference \cite{kob} that it
could be induced by a similarity of the geometry of phase space of these systems
in spite of differences in microscopic dynamics.

Finally, we would like to point out that Fig. \ref{correlation} shows also
the violation of the dissipation-fluctuation theorem (FDT). It seems to be a characteristic
of non-equilibrium system as observed numerically in domain growth process \cite{abarrat},
Lennard-Jones glasses \cite{jlbarrat} or slow granular rheology \cite{berthier} and experimentally
for dielectric measurements
in colloidal glasses \cite{bellon1}, supercooled fluid \cite{grigera} and polymer glasses 
\cite{herisson,buisson}.
Let us consider the response R(t$_w$+t,t$_w$) to a field h conjugated to observable A.
At equilibrium, response to this external field is given by
FDT i.e., at equilibrium, the following equation is verified:

\begin{eqnarray}
        \mathrm{R(t_w+t,t_w)}&\equiv&
\mathrm{\left. \frac{\delta\langle A(t_w+t)\rangle}{\delta h(t_w)} \right|_{h=0}}
\nonumber\\
&=&\mathrm{-\frac{1}{T} \frac {\partial C(t_w+t,t_w)}{\partial t}}
\end{eqnarray}
and then:
\begin{eqnarray}
        \mathrm{\frac {\partial C(t_w+t,t_w)}{\partial t}}&=&
\mathrm{-T\ R(t_w+t,t_w)}
\end{eqnarray}
where T is the bath temperature.

If we consider a quench at a zero-temperature bath
(as in our case), then for satisfying FDT,
normalized correlation function has to be constant:
\begin{equation}
\mathrm{C(t_w+t,t_w)=C(t_w,t_w)=1\ \ ,\ \ \forall t\ge0.}
\end{equation}
Evolutions of correlation functions reported in Fig.~\ref{correlation} show
clearly that C(t$_w$+t,t$_w$) is not a constant with time t: fluctuation-dissipation
theorem is violated in this non-equilibrium system as seen in other glasses 
\cite{abarrat,jlbarrat,berthier,bellon1,grigera,herisson,buisson}. Moreover, the
larger the waiting time, the longer it takes the system to violate the FDT.

\section{Conclusion}

Using simple rheological considerations and results from Burger's
model, we built a phenomenological model of a polymer-chain. This
system is a one dimensional nonlinear lattice characterized by
dissipative couplings. The energy relaxation studies of this
thermalized system show that for sufficiently large viscous
parameter $\gamma$, it is possible to observe nonequilibrium
quasi-stationary states in spite of the short characteristic time
of phonon dissipation! This surprising behavior is due to a chain
auto-organisation which minimizes energy
dissipation, inducing the clamping of some degrees of freedom and
forming long-lived pinned breathers.

Moreover, this very slow energy relaxation can be fitted by
stretched exponential laws, ubiquitous in glassy polymer aging
properties. Another similarity with these physical systems is
that this aging phenomenon is slower when the viscous parameter
$\gamma$ is higher. Furthermore, the two-time correlation
function C(t$_w$+t,t$_w$) shows a strong dependence on the waiting time
and can be scaled onto a master curve by considering the evolution versus
normalized time t/t$_w$.

By these aspects (clamping of degree of
freedom, long-lived nonequilibrium state, stretched exponential
decays, violation of time-translation invariance and master curve), 
the glassy nature of this simple dissipative non linear
lattice is very similar to those observed in glassy polymers.
Beside its interest for nonlinear physics, this model is
presumably an alternative to study complex systems like glassy
state polymers: we have now to push our investigations further by
examining other properties like dependences on cycling
temperature, evolution with waiting time of the shear moduli. Work
along this line is in progress.


\begin{thebibliography}{99}

\bibitem{ediger} M. D. Ediger, C. A. Angell, S. R. Nagel, J. Phys. Chem. {\bf 388}, 13200 (1996)
\bibitem{struck} L. C. E. Struik, {\em Physical Aging in Amorphous Polymers and Other Materials} (Elsevier, Houston, 1978)
\bibitem{young} A. P. Young (Editor), {\em Spin Glasses and Random Fields, Series on Direction
in Condensed Matter Physics}, {\bf 12}, World Scientific,
Singapore (1998)
\bibitem{bellon} L. Bellon, S. Ciliberto, and C. Laroche, Europhys. Lett. {\bf 51}, 551 (2000)
\bibitem{conf} See for examples International Conferences on Relaxation in Complex Systems: Heraklion, 1990,
Proceedings published in J. Non-Cryst. Solids {\bf 131-133} (1991) 1-1266; Alicante, 1993, Proceedings
published in J. Non-Cryst. Solids {\bf 170} (1994) 1-1440
\bibitem{ferry}J. D. Ferry, {\em Viscoelastic Properties of Polymers} (J.Wiley \& Sons, New-York, 1980)
\bibitem{eirich} F. R. Eirich, {\em Rheology} (Academic Press Inc., New-York, 1956)
\bibitem{turner} Turner Alfrey Jr, {\em Mechanical behavior of high polymers} 
(Interscience publishers, Inc., New York, 1948)
\bibitem{aubry} G. P. Tsironis, S. Aubry, Phys. Rev. Lett. {\bf 77}, 5225 (1996)
\bibitem{reigada} R. Reigada, A. Sarmiento, K. Lindenberg, Phys. Rev. E {\bf 64}, 066608 (2001)
\bibitem{peyrard} M. Peyrard and J. Farago, Physica A {\bf 288}, 199 (2000)
\bibitem{kopidakis} G. Kopidakis and S. Aubry, Physica B {\bf 296}, 237 (2001)
\bibitem{aubry2} S. Aubry, K. Kopidakis, e-print cond-mat/0102162
\bibitem{marin} J. L. Marin, F. Falo, P. J. Martinez and L. M. Floria, Phys. Rev. E {\bf 63}, 066603 (2001)
\bibitem{dauxois} T. Dauxois, M. Peyrard and A. R. Bishop, Phys. Rev. E {\bf 47}, 684 (1993)
\bibitem{nose} G. J. Martyna, M. L. Klein, M. Tuckerman, J. Chem. Phys. {\bf 97}, 2635 (1992)
\bibitem{theo} N. Theodorakopoulos, T. Dauxois, M. Peyrard, Phys. Rev. Lett.,  {\bf 85}, 6 (2000)
\bibitem{dau} T. Dauxois, N. Theodorakopoulos, M. Peyrard, J. Stat. Physics, {\bf 107}, 869 (2002)
\bibitem{tsang} K. Y. Tsang, K. L. Ngai, Phys. Rev. E, {\bf 54}, R3067 (1996)
\bibitem{bibaki} A. Bibaki, N. K. Voulgarakis, S. Aubry, G. P. Tsironis, Phys. Rev. E,  {\bf 59}, 1234 (1999)
\bibitem{kohlrausch} R. Kohlrausch, Pogg. Ann. Physik.  {\bf 12}, 393 (1847)
\bibitem{kob} W. Kob and J. L. Barrat, Phys. Rev. Lett. {\bf 78}, 4581 (1997)
\bibitem{vincent} E.Vincent, J. Hammann, M. Ocio, J. P. Bouchaud and L. F. Cugliandolo, 
{\em Slow dynamics and aging}, edited by M. Rubi (Springer-Verlag, Berlin, 1997)
\bibitem{cipelletti} L. Cipelletti, S. Manley, R. C. Ball, D. A. Weitz, Phys. Rev. Lett. {\bf 84}, 2275 (2000)
\bibitem{kroon} M. Kroon, G. H. Wegdam and R. Sprik, Phys. Rev. E {\bf 54}, 6541 (1996)
\bibitem{knaebel} A. Knaebel {\it et al.}, Europhys. Lett. {\bf 52}, 73 (2000)
\bibitem{abarrat} A. Barrat, Phys. Rev. E, {\bf 57}, 3629 (1998)
\bibitem{jlbarrat} J. L. Barrat and W. Kob, Europhys. Lett. {\bf 46}, 637 (1999)
\bibitem{berthier} L. Berthier, L. F. Cugliandolo and J. L. Iguain, Phys. Rev. E, {\bf 63}, 051302 (2001)
\bibitem{bellon1} L. Bellon, S. Ciliberto and C. Laroche, Europhys. Lett. {\bf54}, 511 (2001)
\bibitem{grigera} T. S. Grigera and N. E. Israeloff, Phys. Rev. Lett. {\bf 83}, 5038 (1999)
\bibitem{herisson} D. Herisson and M. Ocio, Phys. Rev. Lett. {\bf 88}, 257202 (2002)
\bibitem{buisson} L. Buisson, S. Ciliberto and A. Garcimartin, in preparation.
\end{thebibliography}
\end{document}